\documentclass[showpacs,preprintnumbers,amsmath,amssymb]{revtex4}
\usepackage{graphicx}
\begin{document}
\preprint{IMAFF-RCA-05-07}
\title{On the accretion of phantom energy onto wormholes}

\author{Pedro F. Gonz\'{a}lez-D\'{\i}az}
\affiliation{Colina de los Chopos, Centro de F\'{\i}sica ``Miguel A.
Catal\'{a}n'', Instituto de Matem\'{a}ticas y F\'{\i}sica
Fundamental,\\ Consejo Superior de Investigaciones Cient\'{\i}ficas,
Serrano 121, 28006 Madrid (SPAIN).}
\date{\today}
\begin{abstract}
By using a properly generalized accretion formalism it is argued
that the accretion of phantom energy onto a wormhole does not make
the size of the wormhole throat to comovingly scale with the scale
factor of the universe, but instead induces an increase of that
size so big that the wormhole can engulf the universe itself
before it reaches the big rip singularity, at least relative to an
asymptotic observer.
\end{abstract}


\maketitle

\vspace{2cm}


\pagebreak

Some attention has been recently paid to the consideration of
cosmic systems formed by wormholes and phantom energy [1,2]. In
particular, it appears of great interest the study of dark energy
accretion onto gravitating systems, such as black holes [3],
wiggly cosmic strings [4] and wormholes [5,6]. In the latter case
it was first claimed [5] that accretion of phantom energy would
induce an increase of the wormhole throat radius so quick that the
wormhole would engulf the entire universe before this reached the
big rip singularity. Such a result has been dubbed big trip and
was later criticized by Faraoni and Israel [6] who have used
several cases of exotic-matter shell on the throat, always
obtaining the conclusion that the wormhole becomes asymptotically
comoving with the cosmic fluid so that the future evolution of the
universe keeps being causal. In this letter I will argue that the
result derived by Faraoni and Israel cannot be recovered if one
uses a proper accretion theory for dark energy onto wormholes.
Actually, the result of Ref. [6] would just describe what one
should expect when the inflationary effects of the accelerated
expansion of the universe on the wormhole size are taken into
account. In fact, when just such effects are considered it is
obtained that wormholes inflate so that their size would scale
like the scale factor of the universe [7], similarly to what later
was obtained by Faraoni and Israel [6].

A proper dark-energy accretion model for wormholes should be
obtained by generalizing the Michel theory [8] to the case of
wormholes. Such a generalization has been already performed by
Babichev, Dokuchaev and Eroshenko [3] for the case of dark-energy
accretion onto Schwarzschild black holes. Here I will adapt the
same procedure to the dark-energy accretion onto the Morris-Thorne
wormhole case. The most general static space-time metric of one
such wormholes is given by [9]
\begin{equation}
ds^2=-e^{\Phi(r)}dt^2 + \frac{dr^2}{1-\frac{K(r)}{r}} +r^2
d\Omega_2^2,
\end{equation}
where the shift function $\Phi(r)$ can be taken to be either zero
or a given function of the radial coordinate $r$, the shape
function $K(r)$ can be taken either as $K(r)=K_0^2/r$ for
wormholes with zero tidal force, or as $K(r)=K_0[1-(r-K_0)/R_0]^2$
if the exotic matter is confined into an arbitrarily small region
with width $R_0$ around the wormhole throat, and $d\Omega_2^2$ is
the metric on the unit two-sphere.

We shall follow now the procedure used by Babichev, Dokuchaev and
Eroshenko [3], adapting it to the case of a Morris-Thorne
wormhole, and thus consider the tensor momentum-energy for a
perfect fluid
\begin{equation}
T_{\mu\nu}=(p+\rho)u_{\mu}u_{\nu}+pg_{\mu\nu},
\end{equation}
in which $p$ is pressure, $\rho$ is energy density and
$u^{\mu}=dx^{\mu}/ds$ is the four-velocity. By integrating then the
time component of the conservation law for momentum-energy tensor,
$T_{\mu;\nu}^{\nu}=0$, we can obtain for the spherical symmetry of
our metric,
\begin{equation}
u m^{-2} r^2\left(1-\frac{K(r)}{r}\right)^{-1}(p+\rho)\sqrt{u^2
+\frac{K(r)}{r}-1}=C ,
\end{equation}
where $m$ is the exotic mass of the wormhole which, following the
procedure of Ref. [3], has been introduced to render the
integration constant $C$ to have the dimensions of an energy
density (note that we are using natural units so that
$G=c=\hbar=1$), $u=dr/ds$ and, without any loss of generality for
our present purposes, we have adhered to the case where $\Phi=0$.
We now also integrate the projection of the conservation law for
momentum-energy tensor onto the four-velocity,
$u_{\mu}T^{\mu\nu}_{;\nu}=0$. For a perfect fluid and spherical
symmetry, since $u>0$, we finally obtain
\begin{equation}
m^{-2} r^2 u \left(1-\frac{K(r)}{r}\right)^{-1/2}
e^{\int_{\rho_{\infty}}^{\rho}\frac{d\rho}{p+\rho}} =A ,
\end{equation}
with $A$ a positive dimensionless integration constant. From Eqs.
(3) and (4) we get finally
\begin{equation}
\left(1-\frac{K(r)}{r}\right)^{-1/2}
\sqrt{u^2+\frac{K(r)}{r}-1}(p+\rho)e^{-\int_{\rho_{\infty}}^{\rho}\frac{d\rho}{p+\rho}}
= B,
\end{equation}
in which $B=C/A=
\hat{A}\left(\rho_{\infty}+p(\rho_{\infty})\right)$, with $\hat{A}$
a positive constant.

The most general expression for the rate of change of the exotic
wormhole mass should be given by integrating the momentum density
$T_0^r$ over the element of two dimensional spherical surface
$dS=r^2 \sin\theta d\theta d\phi$ [10], i.e.
\begin{equation}
\dot{m}=\int dS T_0^r,
\end{equation}
where the sign has been chosen so that the rate refers to a negative
energy. We then obtain from Eqs. (4) and (5),
\begin{equation}
\dot{m}=-4\pi m^2 Q\sqrt{1-\frac{K(r)}{r}}(p+\rho) ,
\end{equation}
with the constant $Q=A\hat{A}>0$. For the relevant asymptotic
regime $r\rightarrow\infty$, the rate $\dot{m}$ reduces to
\begin{equation}
\dot{m}=-4\pi m^2 Q(p+\rho) .
\end{equation}
We see then that the rate for the wormhole exotic mass due to
accretion of dark energy becomes exactly the negative to the
similar rate in the case of a Schwarzschild black hole,
asymptotically. It should be pointed out however that it is not a
necessary condition that the wormhole possesses negative energy
densities (as measured by static observers), but that it violates
the null energy condition. Had we considered a positive sign in
Eq. (7) and a positive wormhole energy density (and positive ADM
mass), the respective accretion of phantom energy had been
accompanied with a diminishing of the wormhole mass, as in the
case of black holes analyzed by Babichev, Dokuchaev and Eroshenko
in Ref. [3].

For a quintessence model with equation of state
$p=w\rho=w\rho_0a^{-3(1+w)}$ and scale factor
\begin{equation}
a=T^{2/[3(1+w)]}=
\left(a_0^{3(1+w)/2}+\frac{3}{2}(1+w)C(t-t_0)\right)^{2/[3(1+w)]}
,
\end{equation}
with $C=\sqrt{8\pi G\rho_0/3}$, one can integrate Eqs. (7) and (8).
For the phantom regime $w<-1$, we obtain from the asymptotic
expression (8) the evolution of the exotic mass of the wormhole,
\begin{equation}
m=\frac{m_0}{1-\frac{4\pi Q\rho_0 m_0(|w|-1)(t-t_0)}{T(w<-1)}} ,
\end{equation}
where $m_0$ is the initial exotic mass of the wormhole. This
result is consistent with what was obtained in Ref. [5], implying
that the exotic mass $m$ diverges at the time
\begin{equation}
t_*=t_0+\frac{t_{br}-t_0}{1+\frac{8\pi Q m_0 a_0^{3(|w|-1)/2}}{3C}}
,
\end{equation}
where $t_{br}$ is the time at which the big rip takes place, i.e.
$t_{br}=t_0+2a_0^{-3(|w|-1)/2}/[3(|w|-1)C]$. Thus, $t_*$ occurs
before the big rip. Now, depending on the distribution of the exotic
mass on the throat, we can derive the variation of the wormhole
throat radius with phantom energy accretion for the asymptotic case.
In particular, for a spherical shell distribution with constant
thickness, the wormhole throat radius $K_0$will be given by
\begin{equation}
K_0=\frac{K_{0i}}{\sqrt{1-\frac{4\pi Q\rho_0
m_0(|w|-1)(t-t_0)}{T(w<-1)}}} ,
\end{equation}
with $K_{0i}$ the initial size of the wormhole throat. In this
case the wormhole can only evolve until the time $t_*$. If we take
$K_0\propto m$, then we would get the same result as derived in
Ref. [5] for the wormhole throat, this time without assuming any
relation between the sizes of the wormhole and the corresponding
Schwarzschild black hole. In any event, if we let
$r\rightarrow\infty$, the size of the wormhole would exceed the
size of the universe before this reached the big rip singularity,
so restoring the problem with causality in the future of the
universe.

Thus, by applying a proper accretion dark energy theory to the
wormhole, we obtain that, superposed to the inflationary effects
that the universal acceleration has on the wormhole size which
thereby scales like the scale factor [6,7], there is a real income
of dark energy which adds to the mass of the wormhole when $w<-1$.
In the asymptotic case $r\rightarrow\infty$, big trips violating
causality could then be performed by the universe as a whole if
the equation-of-state parameter of the universe would keep on a
constant value less than -1 in the future.

The above conclusion is only strictly valid if we allow
$r\rightarrow\infty$. For otherwise, if we assume e.g. $K_0^2=\gamma
m$, with $\gamma$ a constant, for wormholes with zero tidal forces,
the integration of Eq. (7) yields
\begin{equation}
\left.\left[\frac{\sqrt{1-\frac{\gamma m}{r^2}}}{m}
+\frac{\gamma}{2r^2}\ln\left(\frac{2-\frac{\gamma
m}{r^2}+2\sqrt{1-\frac{\gamma
m}{r^2}}}{m}\right)\right]\right|_{m_0}^m =-\frac{4\pi
Q\rho_0(1+w)(t-t_0)}{a_0^{3(1+w)/2}T} .
\end{equation}
It can be checked that the accretion of phantom energy in this case
also induces an increase of the wormhole exotic mass, but that
increase can just be produced until $m$ reaches a maximum value
given by $m\leq m_{max}=r^2/\gamma$.

I consider the conclusions drawn in the present letter to be weird
but still correct. It could be at first sight argued that since
metric (1) is static our analysis can only be a good approximation
for very small accretion rates. However, even though the back
reaction on the wormhole metric (1) has not been explicitly
considered, the general conclusions of our study must still remain
valid at any accretion rate and hence for the big trip case, as
should be Eqs. (3) and (4), because the modifications that the use
of a time-dependent wormhole metric induced in these expressions
are all expected to vanish on the asymptotic limit where the big
trip feature should occur, at least if all time-dependence in the
metric is confined to take place only in the throat radius so that
the shape of the wormhole is preserved during phantom energy
accretion. In this case, the use of a non static wormhole metric
would lead to significant changes in all of our calculations for
$r<\infty$, but leave the calculation given in this letter
essentially unchanged and valid at any accretion rate
asymptotically.

\acknowledgements

\noindent This work was supported by MCYT under Research Project No.
BMF2002-03758.

\end{document}